\documentclass{article}

\usepackage{hyperref}
\usepackage{amsfonts,amsmath, amsthm, amssymb}
\usepackage{graphicx}
\usepackage{color}

\newcommand{\ket}[1]{\lvert #1 \rangle} 
\newcommand{\bra}[1]{\langle #1 \rvert} 
\newcommand{\+}         {\dagger}

\DeclareMathOperator{\tr}{\mathrm{tr}}
\DeclareMathOperator{\eqdef}{\triangleq}

\newcommand{\ff}{{\mathbb F}_4}
\newcommand{\gG}{\mathsf G}
\newcommand{\gH}{\mathsf H}
\newcommand{\gK}{\mathsf K}
\newcommand{\I}{\mathbb I}

\newtheorem{lemma}{Lemma}
\newtheorem{proposition}{Proposition}
\newtheorem{definition}{Definition}
\newtheorem{example}{Example}
\newtheorem{remark}{Remark}

\begin{document}

\title{Constructions and performance of classes of quantum LDPC codes}

\author{Thomas Camara\thanks{INRIA, Projet Codes, BP 105, Domaine de Voluceau, F-78153 Le Chesnay, France.}
\and 
Harold Ollivier\thanks{Perimeter Institute, 31 Caroline St. N, Waterloo, ON N2L 2Y5, Canada.}
\and Jean-Pierre Tillich\thanks{INRIA, Projet Codes, BP 105, Domaine de Voluceau, F-78153 Le Chesnay, France.}}

\date{}

\maketitle

\begin{abstract}
Two methods for constructing quantum LDPC codes are presented. We
explain how to overcome the difficulty of finding a set of low weight
generators for the stabilizer group of the code. Both approaches are
based on some graph representation of the generators of the stabilizer
group and on simple local rules that ensure commutativity. A message
passing algorithm for generic quantum LDPC codes is also
introduced. Finally, we provide two specific examples of quantum LDPC
codes of rate 1/2 obtained by our methods, together with a numerical
simulation of their performance over the depolarizing channel.
\end{abstract}

\section{Introduction}
The idea of using quantum systems for processing information has been
suggested by R.~P.~Feynman as a potential way of bypassing the difficulty of
simulating quantum physics with classical computers. Since then, it
has developed into an exciting research area with implications ranging
from cryptography (see e.g.~\cite{BB84a,Eke91a, SP00a}) to complexity
theory (e.g.~\cite{Amb04a,Reg04a}). For instance, quantum computers
could solve efficiently some hard problems such as integer
factorization~\cite{Sho94a}, or give quadratic speed-up over optimal
classical algorithms, as it is the case for unsorted database
search~\cite{Gro96a}.

However, for taking advantage of the quantum nature of physical
systems to process information, it is necessary to protect them from
unwanted evolutions. Indeed, if quantum registers are not protected
from noise, the very fragile superpositions required for efficiently
manipulating quantum information tend to disappear exponentially fast
with the number of qubits involved. This effect ---~called
decoherence~--- can nonetheless be reduced by using quantum error
correcting codes. The first scheme of this kind has been proposed in
1995 by P.~Shor~\cite{Sho95a}, and triggered numerous work on quantum
error correction. Most notable, was the introduction of the stabilizer
formalism~\cite{Got97a,CRSS98a} ---~for defining quantum codes and
finding the gate implementation of encoding and decoding circuits~---
together with the class of CSS codes~\cite{CS96a,Ste96a}. With these
tools at hand, it has been shown that quantum information processing
can be done fault-tolerantly (see for instance \cite{AB97a}),
i.e. would be feasible even in the presence of qubit errors and gate
faults ---~provided these events are rare enough.

In spite of these important results, properties of quantum codes are
less understood than those of for their classical counterparts. It is
thus of interest to tackle the problem from a pragmatic point of view:
devise versatile constructions of quantum codes inspired by the best
classical codes, and analyze their performances. 

When sending classical information over memoryless classical channels,
it has been demonstrated that a very efficient way for approaching the
channel capacity is obtained by using LDPC codes with Gallager's
iterative decoding algorithm. Generalizing these notions to quantum
codes seems a promising way, and has indeed been proposed
recently~\cite{MMM04a}. In this work, D.\@~J.\@~C.\@~MacKay {\it et
al.}  have shown how to construct sparse weakly self-dual binary codes
that can be used to construct quantum LDPC codes using
Calderbank-Shor-Steane's method.

Our work is aimed at finding other constructions of quantum LDPC codes
within the stabilizer formalism. While a brief introduction to
stabilizer codes is provided below, we would like to pinpoint here the
main difficulty for finding LDPC stabilizer codes. As explained
in~\cite{CRSS98a}, each stabilizer code can be viewed as a code over
$\ff$, the field with four elements. However, the converse is not
true: to correspond to valid stabilizer codes, these codes must be
self-orthogonal for some Hermitian trace inner product. Fulfillment of
this peculiar constraint makes usual constructions of LDPC codes
useless in the quantum setting. Our work provides a partial solution
to this problem by defining such codes through simple group
theoretical constructions (see Sections \ref{sec:generic} and
\ref{sec:4_8}).  The resulting codes can be viewed as quantum analogs
of regular Gallager codes: each qubit is involved in the same number
of parity-check equations, and each parity-check equation involves
the same number of qubits. Error estimation has been performed by an
iterative algorithm applied to a Tanner graph associated to our
constructions. The results of these numerical simulations are
presented in Section \ref{sec:results}.

\section{Stabilizer codes}
\paragraph{The code subspace.}
An $(n,k)$ quantum code is a subspace of dimension $2^k$ of the
Hilbert space $\mathcal H_n \cong (\mathbb C^2)^{\otimes n}$ of $n$
qubits.  Such subspace allows to encode $k$ qubits, so that the rate
of an $(n,k)$ quantum code is $\frac{k}{n}$.  While defining the code
subspace can be done in many different ways, a particularly useful
method is known as the stabilizer formalism.

The definition of stabilizer codes relies heavily on properties of Pauli
matrices for one qubit, i.e. acting on $\mathcal H_1$:
\begin{equation*}
I=\left(\begin{array}{cc}1&0\\0&1\end{array}\right), 
X= \left(\begin{array}{cc}0&1\\1&0\end{array}\right),  
Y= \left(\begin{array}{cc}0&-i\\i&0\end{array}\right), 
Z= \left(\begin{array}{cc}1&0\\0&-1\end{array}\right). 
\end{equation*}
In particular, these matrices square to the identity; they generate a
finite multiplicative group $\mathcal G_1$; and any pair $(P,Q)$ of
elements in $\mathcal G_1$ either commutes ($ PQ - QP = 0$) or
anti-commutes ($PQ + QP = 0$).

By extension, the Pauli group $\mathcal G_n$ over $n$-qubits is the
multiplicative group generated by all possible tensor products of
Pauli matrices over $n$ qubits, i.e. elements of the form $P_1\otimes
P_2\otimes \ldots \otimes P_n$ where the $P_i$'s belong to $\mathcal
G_1$. As in the case of the Pauli group of a single qubit, any pair of
elements in $\mathcal G_n$ either commutes or anti-commutes.

The code subspace $\mathcal C$ of a $(n,k)$ stabilizer code is defined
as the largest subspace of $\mathcal H_n$ stabilized by the action of
a subgroup $\mathcal S$ of $\mathcal G_n$. For $\mathcal C$ to be of
dimension $2^k$, it requires $\mathcal S$ to be an Abelian subgroup
generated by $n-k$ independent elements $M_i$, and such that
$-I^{\otimes n} \notin \mathcal S$. That is, the code subspace is
defined as,
$$\ket \psi \in \mathcal C \Leftrightarrow \forall i, \ M_i \ket \psi
= \ket \psi.$$

\paragraph{Error estimation.}
Before explaining how errors can be estimated, we must first define
what the possible errors are. It is well known that the most general
transformation on $n$ qubits allowed by quantum mechanics is a
completely positive trace preserving map. However, as shown in
\cite{BDSW96a, EM96a, KL97a}, it is sufficient to consider errors as
elements of $\mathcal G_n$ since any admissible quantum operation can
be written in Kraus form using only elements of the Pauli group over
$n$ qubits.

Thus, let $E \in \mathcal G_n$ be an error on the qubits of a $(n,k)$
stabilizer code with stabilizer group $\mathcal S = \langle M_i
\rangle_{i=1}^{n-k}$. For each $i$, $E$ and $M_i$ either commute or
anti-commute. Hence, for $\ket \psi \in \mathcal C$, either $M_i E\ket
\psi = E\ket \psi$ or $M_i E\ket \psi = -E \ket \psi$. The action of
$E$ on $\mathcal C$ is either to map $\mathcal C$ onto itself, or to
map $\mathcal C$ onto an orthogonal subspace of $\mathcal H_n$. Thus,
the measurement associated to the $M_i$'s reveals onto which
orthogonal subspace $\mathcal C$ has been mapped: the $n-k$
dimensional binary vector $s(E) = (s_1(E), \ldots, s_{n-k}(E))$, such
that $(-1)^{s_i(E)} = \bra \psi E^\+ M_i E \ket \psi$, characterizes
$E\mathcal C$, and is called the syndrome of $E$.

Also note that not all errors $E$ with $s(E) = (0,\ldots,0)$ are
harmful. Indeed if $E \in \mathcal S$, then $\forall \ket \psi \in
\mathcal C$, we have $E \ket \psi = \ket \psi$, and quantum
information is preserved. Only errors with non-trivial action on
$\mathcal C$ harm quantum information. These errors $E$ are by
definition elements of $N(\mathcal S) - \mathcal S$, where $N(\mathcal
S)$ is the normalizer of $\mathcal S$.

A generic error estimation procedure then consists in finding $\hat E
\in \mathcal G_n$ given the error model for the $n$ qubits and the
syndrome $s(E)$ such that $\hat E^\+ E \in \mathcal S$. Such $\hat E$
is called the estimated error. The most common error model considered
in the literature is the depolarizing channel where $X$, $Y$ and $Z$
errors occur each with probability $p/3 < 1/4$, and independently on
each qubit. Usually, the estimated error is chosen accordingly to the
maximum likelihood criterion which is equivalent to choosing for $\hat
E$ the lowest weight element of $\mathcal G_n$ such that $s(\hat E) =
s(E)$. The error recovery will be successful when $\hat E^\+ E\in
\mathcal S$, otherwise a block error is produced.

\paragraph{Quantum codes as codes over $\ff$.}
Stabilizer codes might also be viewed as codes over $\ff =
\{0,1,\omega, \bar\omega\}$, with $1+\omega+\omega^2 =
1+\omega+\bar\omega=0$ . This duality is due to the additive structure
of $\ff$ which echoes the multiplicative structure of the Pauli
operators $I,X,Y,Z$. The mapping between one and the other is the
following: $I\leftrightarrow 0$, $X \leftrightarrow \omega$, $Z
\leftrightarrow \bar \omega$ and $Y \leftrightarrow 1$. In addition,
with the trace operation $\tr(x) \eqdef x + \bar x = x+x^2$, it can be
checked that $\tr(a \bar b)$ is $0$ iff the Pauli operators associated
to $a$ and $b$ commute and $1$ iff they anti-commute. This motivates
the definition of the following inner product over $\ff^n$ :
$$<u,v> \eqdef \tr \sum_{i=1}^n u_i \bar v_i.$$

The generators of a stabilizer code, when expressed as row-vectors of
$\ff^n$, form a matrix $M$ which will be called a parity-check matrix
of the stabilizer code.  In what follows, both the $i$-th generator
defining the stabilizer code and its expression as the $i$-th row of
$M$ are denoted by $M_i$. These rows are orthogonal with respect to
the inner product defined above.  The converse is also true: any
$(n-k) \times n$ matrix over $\ff$ with orthogonal rows defines a
stabilizer code over $n$ qubits.

While stabilizer codes were originally introduced using the language
of the Pauli group, we will mainly employ the $\ff$ formalism. This
choice simplifies the notation, and better points out analogies and
differences in constructing classical and quantum LDPC
codes. Interested readers are redirected to \cite{Got97a, CRSS98a} for
a more complete treatment of stabilizer codes such as gate
implementation and additional connections between stabilizer codes and
codes over $\ff$.

\section{Constructing regular quantum LDPC codes}

According to the definition of a parity-check matrix for stabilizer
codes, it is natural to introduce quantum LDPC codes as stabilizer
codes with sparse parity-check matrix:

\begin{definition}
An $(a,b)$-regular quantum LDPC code is a stabilizer code whose
parity-check matrix $M$ has ``$a$'' non-zero entries per column and
``$b$'' non-zero entries per row.
\end{definition}

In the rest of this article we are going to construct and analyze some
$(a,b)$-regular LDPC codes.

\subsection{Iterative error estimation for quantum LDPC codes}
\label{sec:iterative}

To define the iterative error estimation algorithm we use the
following Tanner graph associated to a parity check matrix $M$. It is
a bipartite graph with vertex set $V\cup W$. The vertices of $V$ are
associated to qubits of the code (i.e. columns of $M$) and vertices of
$W$ to the generators of the stabilizer code (i.e. rows of $M$).
There is an edge between $v \in V$ and $w \in W$ iff the corresponding
entry $M_{wv}$ in $M$ is non zero. We label this edge with the entry
$M_{wv}$. The purpose of the iterative error estimation algorithm is
to find the most likely value for each $\hat E_i$ such that $\hat E =
(\hat E_1, \ldots, \hat E_n)$ is compatible with the observed syndrome
$s(E) = (s_i(E))_{i=1}^{n-k}$.  The local constraint on $\hat E$
associated to the $j$-th check node (i.e. vertices in $W$) is given by
the equation $<\hat E,M_j> = s_j(E).$

This graph is nothing but the syndrome version of a Tanner graph. The
algorithms we can use to estimate errors are the SUM-PRODUCT or the
MIN-SUM algorithms in their syndrome decoding form applied to this
graph. That is, the version of these algorithms which estimates the
most likely value of the error for each qubit given the observed
syndrome. Note that iterative decoding of classical codes estimates
instead the most likely value of the codeword symbols for each bit of
the code.

\subsection{Generic construction of $(a,b)$-regular quantum LDPC codes}
\label{sec:generic}

In order to construct quantum LDPC codes we restrict our attention to
codes having a parity check matrix $M$ with two kinds of non zero
entries, $\omega$ and $\bar \omega$.

We start our construction by choosing a group $\gG$ with cardinality
equal to a multiple of the length of the code we are interested
in. Then we choose two subgroups $\gH$ and $\gK$ of $\gG$, with $|\gK|
> |\gH|$. The cosets $x\gH$ are associated to qubits whereas the
cosets $y\gK$ are associated to rows of the parity-check matrix. In
other words the length $n$ and the number of rows $n-k$ in the parity
check matrix are given by
$$n = \frac{|\gG|}{|\gH|}, \quad n-k = \frac{|\gG|}{|\gK|}.$$ 
We then pick a set of generators $G$ of $\gG$ that can be partitioned
into two sets $G = G_\omega \cup G_{\bar\omega}$ such that the following
properties are satisfied:
\begin{eqnarray}
& & (G_\omega)^{-1} = G_\omega, \quad (G_{\bar\omega})^{-1} =
G_{\bar\omega} \label{eq:commutation1} \\ & &
\forall(g_\omega,g_{\bar\omega}) \in G_\omega \times G_{\bar\omega},
\quad g_\omega g_{\bar\omega} = g_{\bar\omega} g_\omega
\label{eq:commutation3}\\ & &
g,g'\in G,\ h,h'\in\gH,\ k,k'\in\gK,\
ghk=g'h'k'\ \implies\ g=g' \label{eq:simple_edge}
\end{eqnarray}
 
We put an edge between coset $x\gH$ and coset $y\gK$ iff there exists
$g \in G$ such that $xg\gH \cap y\gK \neq \emptyset$, or equivalently
iff there exist $h \in \gH$, $k \in \gK$ such that $y=xghk$.  We label
this edge with a $\omega$ if the corresponding $g$ belongs to
$G_\omega$ and with a $\bar\omega$ otherwise. It can be checked that
the degree of any vertex $x\gH$ is equal to $a \eqdef {|G| |\gH|} /
{|\gH \cap \gK|}$ and the degree of any vertex $y\gK$ is equal to $b
\eqdef {|G| |\gK|} / {|\gH \cap \gK|}$. This is a simple consequence
of the following lemma:
\begin{lemma}
\label{lem:easy}
The intersection of a coset $x\gH$ and $y\gK$ is either empty or equal to
a coset $z(\gH \cap \gK)$.
\end{lemma}

This defines the Tanner graph of our code and therefore also its
parity-check matrix $M$. Property (\ref{eq:simple_edge}) imposes that
there are no multiple edges in the graph. The point of this
construction is that the commutation of the $g_\omega$'s with the
$g_{\bar\omega}$'s implies the orthogonality of the rows of $M$. Thus,
the matrix $M$ is a valid parity-check matrix for defining a
stabilizer code.

\begin{proposition}
The parity-check matrix $M$ associated to the Tanner graph given by
this construction has orthogonal rows.
\end{proposition}

\begin{proof}
Given two rows $M_i$ and $M_j$ of $M$, a parity-check matrix of a
stabilizer code, one can partition the set of qubits in two classes:
qubits for which the corresponding entries in $M_i$ and $M_j$ have
inner product $0$, and those for which the entries have inner product
$1$. For $M_i$ and $M_j$ to be orthogonal, it is necessary and
sufficient that the number of qubits of the second type is even.
Equivalently, and this is how our construction is tailored, we must
show that qubits of the second type can be paired together.

Consider a qubit belonging to the second class; say it corresponds to
the vertex $x\gH$ and the rows $M_i$ and $M_j$ correspond to the
cosets $y\gK$ and $z\gK$. Since for this qubit the entries of $M_i$
and $M_j$ have inner product $1$, one of these two rows has a $\omega$
at the position of $x\gH$ and the other one has a $\bar \omega$.
Without loss of generality we may assume that the subgraph of the
Tanner graph induced by $x\gH$, $y\gK$ and $z\gK$ is as in
Fig.~\ref{fig:subgraph}. In other words there exist $g_\omega \in
G_\omega$, $g_{\bar\omega} \in G_{\bar\omega}$ such that $xg_\omega\gH
\cap y\gK \neq \emptyset$ and $xg_{\bar\omega} \gH \cap z\gK \neq
\emptyset$.  Let $x' \eqdef x g_\omega g_{\bar\omega}$. Note that there exist
$h_1,h_2$ in $\gH$ and $k_1,k_2$ in $\gK$ such that
\begin{eqnarray}
y & = & xg_\omega h_1k_1 \label{eq:first_edge} \\
z & = & xg_{\bar\omega}h_2k_2 \label{eq:second_edge}
\end{eqnarray}
Since $xg_{\omega}=x'g_{\bar\omega}^{-1}$, using (\ref{eq:first_edge})
implies that $y=x'g_{\bar\omega}^{-1}h_1k_1$. Therefore, there is an
edge labeled by $\bar\omega$ between $x'\gH$ and $y\gK$. Because
$g_\omega g_{\bar\omega}=g_{\bar\omega} g_\omega$ we also have
$xg_{\bar\omega} = x'g_\omega^{-1}$. A similar reasoning shows that $z
= x'g_\omega^{-1}h_2k_2$, which implies that there is an edge labeled
by $\omega$ between $x'\gH$ and $z\gK$. Each qubit of the second class
is necessarily involved in a $4$-cycle with another such qubit (See
Fig.~\ref{fig:4cycle}).

\begin{figure}
\begin{center}
\setlength{\unitlength}{2072sp}%
\begingroup\makeatletter\ifx\SetFigFont\undefined%
\gdef\SetFigFont#1#2#3#4#5{%
  \reset@font\fontsize{#1}{#2pt}%
  \fontfamily{#3}\fontseries{#4}\fontshape{#5}%
  \selectfont}%
\fi\endgroup%
\begin{picture}(9384,2094)(1801,-3358)
\put(10801,-1456){\makebox(0,0)[lb]{\smash{{\SetFigFont{10}{12.0}{\familydefault}{\mddefault}{\updefault}{\color[rgb]{0,0,0}$g_\omega$}%
}}}}
\thinlines
{\color[rgb]{0,0,0}\put(8101,-2176){\line( 0,-1){270}}
}%
{\color[rgb]{0,0,0}\put(7966,-2311){\line( 1, 0){270}}
}%
{\color[rgb]{0,0,0}\put(6436,-1411){\circle{270}}
}%
{\color[rgb]{0,0,0}\put(6436,-1276){\line( 0,-1){270}}
}%
{\color[rgb]{0,0,0}\put(6301,-1411){\line( 1, 0){270}}
}%
{\color[rgb]{0,0,0}\put(6436,-3211){\circle{270}}
}%
{\color[rgb]{0,0,0}\put(6436,-3076){\line( 0,-1){270}}
}%
{\color[rgb]{0,0,0}\put(6301,-3211){\line( 1, 0){270}}
}%
{\color[rgb]{0,0,0}\put(8101,-1411){\circle{270}}
}%
{\color[rgb]{0,0,0}\put(2566,-1411){\circle{270}}
}%
\thicklines
{\color[rgb]{0,0,0}\put(2701,-1411){\line( 1, 0){3600}}
}%
{\color[rgb]{0,0,0}\put(10126,-1411){\line( 1, 0){450}}
}%
\thinlines
{\color[rgb]{0,0,0}\put(2701,-1411){\line( 2,-1){3600}}
}%
{\color[rgb]{0,0,0}\put(2701,-1456){\line( 2,-1){3600}}
}%
{\color[rgb]{0,0,0}\put(10126,-2311){\line( 1, 0){450}}
}%
{\color[rgb]{0,0,0}\put(10126,-2356){\line( 1, 0){450}}
}%
\put(8551,-2356){\makebox(0,0)[lb]{\smash{{\SetFigFont{10}{12.0}{\familydefault}{\mddefault}{\updefault}{\color[rgb]{0,0,0}check-node}%
}}}}
\put(8551,-1456){\makebox(0,0)[lb]{\smash{{\SetFigFont{10}{12.0}{\familydefault}{\mddefault}{\updefault}{\color[rgb]{0,0,0}qubit}%
}}}}
\put(6751,-1456){\makebox(0,0)[lb]{\smash{{\SetFigFont{10}{12.0}{\familydefault}{\mddefault}{\updefault}{\color[rgb]{0,0,0}$y\gK$}%
}}}}
\put(1801,-1456){\makebox(0,0)[lb]{\smash{{\SetFigFont{10}{12.0}{\familydefault}{\mddefault}{\updefault}{\color[rgb]{0,0,0}$x\gH$}%
}}}}
\put(6751,-3256){\makebox(0,0)[lb]{\smash{{\SetFigFont{10}{12.0}{\familydefault}{\mddefault}{\updefault}{\color[rgb]{0,0,0}$z\gK$}%
}}}}
\put(10801,-2356){\makebox(0,0)[lb]{\smash{{\SetFigFont{10}{12.0}{\familydefault}{\mddefault}{\updefault}{\color[rgb]{0,0,0}$g_{\bar\omega}$}%
}}}}
{\color[rgb]{0,0,0}\put(8101,-2311){\circle{270}}
}%
\end{picture}%

\end{center}
\caption{Subgraph induced by the qubit $x\gH$ and the two check-nodes
$y\gK$ and $z\gK$.}\label{fig:subgraph}
\end{figure}
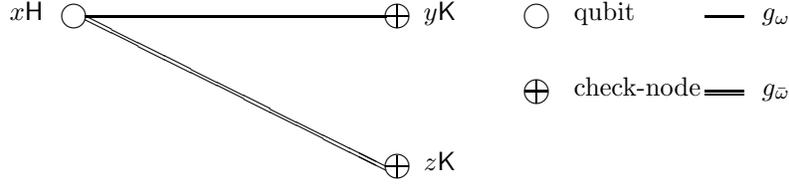

\begin{figure}
\begin{center}
\setlength{\unitlength}{2072sp}%
\begingroup\makeatletter\ifx\SetFigFont\undefined%
\gdef\SetFigFont#1#2#3#4#5{%
  \reset@font\fontsize{#1}{#2pt}%
  \fontfamily{#3}\fontseries{#4}\fontshape{#5}%
  \selectfont}%
\fi\endgroup%
\begin{picture}(9384,2094)(1801,-3358)
\put(1801,-3256){\makebox(0,0)[lb]{\smash{{\SetFigFont{10}{12.0}{\familydefault}{\mddefault}{\updefault}{\color[rgb]{0,0,0}$x'\gH$}%
}}}}
\thinlines
{\color[rgb]{0,0,0}\put(8101,-2176){\line( 0,-1){270}}
}%
{\color[rgb]{0,0,0}\put(7966,-2311){\line( 1, 0){270}}
}%
{\color[rgb]{0,0,0}\put(6436,-1411){\circle{270}}
}%
{\color[rgb]{0,0,0}\put(6436,-1276){\line( 0,-1){270}}
}%
{\color[rgb]{0,0,0}\put(6301,-1411){\line( 1, 0){270}}
}%
{\color[rgb]{0,0,0}\put(6436,-3211){\circle{270}}
}%
{\color[rgb]{0,0,0}\put(6436,-3076){\line( 0,-1){270}}
}%
{\color[rgb]{0,0,0}\put(6301,-3211){\line( 1, 0){270}}
}%
\thicklines
{\color[rgb]{0,0,0}\put(10126,-1411){\line( 1, 0){450}}
}%
\thinlines
{\color[rgb]{0,0,0}\put(10126,-2311){\line( 1, 0){450}}
}%
{\color[rgb]{0,0,0}\put(10126,-2356){\line( 1, 0){450}}
}%
\put(10801,-1456){\makebox(0,0)[lb]{\smash{{\SetFigFont{10}{12.0}{\familydefault}{\mddefault}{\updefault}{\color[rgb]{0,0,0}$g_\omega$}%
}}}}
\put(10801,-2356){\makebox(0,0)[lb]{\smash{{\SetFigFont{10}{12.0}{\familydefault}{\mddefault}{\updefault}{\color[rgb]{0,0,0}$g_{\bar\omega}$}%
}}}}
{\color[rgb]{0,0,0}\put(8101,-1411){\circle{270}}
}%
{\color[rgb]{0,0,0}\put(2566,-3211){\circle{270}}
}%
{\color[rgb]{0,0,0}\put(2566,-1411){\circle{270}}
}%
\thicklines
{\color[rgb]{0,0,0}\put(2701,-3211){\line( 1, 0){3600}}
}%
\thinlines
{\color[rgb]{0,0,0}\put(2701,-3211){\line( 2, 1){3600}}
}%
{\color[rgb]{0,0,0}\put(2701,-1411){\line( 2,-1){3600}}
}%
\thicklines
{\color[rgb]{0,0,0}\put(2701,-1411){\line( 1, 0){3600}}
}%
\thinlines
{\color[rgb]{0,0,0}\put(2701,-1456){\line( 2,-1){3600}}
}%
{\color[rgb]{0,0,0}\put(2701,-3166){\line( 2, 1){3600}}
}%
\put(8551,-2356){\makebox(0,0)[lb]{\smash{{\SetFigFont{10}{12.0}{\familydefault}{\mddefault}{\updefault}{\color[rgb]{0,0,0}check-node}%
}}}}
\put(8551,-1456){\makebox(0,0)[lb]{\smash{{\SetFigFont{10}{12.0}{\familydefault}{\mddefault}{\updefault}{\color[rgb]{0,0,0}qubit}%
}}}}
\put(6751,-1456){\makebox(0,0)[lb]{\smash{{\SetFigFont{10}{12.0}{\familydefault}{\mddefault}{\updefault}{\color[rgb]{0,0,0}$y\gK$}%
}}}}
\put(1801,-1456){\makebox(0,0)[lb]{\smash{{\SetFigFont{10}{12.0}{\familydefault}{\mddefault}{\updefault}{\color[rgb]{0,0,0}$x\gH$}%
}}}}
\put(6751,-3256){\makebox(0,0)[lb]{\smash{{\SetFigFont{10}{12.0}{\familydefault}{\mddefault}{\updefault}{\color[rgb]{0,0,0}$z\gK$}%
}}}}
{\color[rgb]{0,0,0}\put(8101,-2311){\circle{270}}
}%
\end{picture}%

\end{center}
\caption{Subgraph showing that each qubit $x\gH$ of the second type
(involved in $M_i$ with a $\omega$ and in $M_j$ with a $\bar\omega$)
is necessarily part of a 4-cycle with another qubit $x'\gH = xg_\omega
g_{\bar\omega}\gH$ of the second type. }\label{fig:4cycle}
\end{figure}
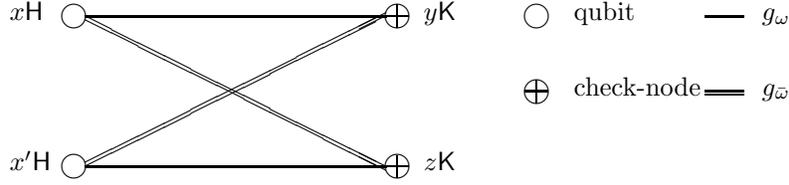

From the absence of multiple edges in our Tanner graph, qubits of the
second class can always be arranged in pairs forming the $4$-cycle
described above. Hence, $<M_i,M_j>=0$ for any $i$ and $j$, and $M$ is
the parity-check matrix of a stabilizer code.
\end{proof}

\begin{remark} \rm
For satisfying the commutation constraints, it is necessary to
introduce many $4$-cycles in the Tanner graph associated to stabilizer
codes. More precisely, let us define the {\em 4-cycle graph}
associated to a Tanner graph as the graph with vertex set the qubits
and with edges connecting two qubits iff they are involved in the same
$4$-cycle in the Tanner graph. The 4-cycle graph associated to our
construction is a regular graph with degree at least
${|G_\omega||G_{\bar\omega}| |\gH|^2}/{|\gH \cap \gK|^2} $.  More
generally, $(a,b)$-regular codes, where each qubit is involved in
$a_\omega$ generators with a $\omega$ and $a_{\bar\omega}$ generators
with a $\bar\omega$, have a $4$-cycle graph where each vertex has
degree at least $a_\omega a_{\bar\omega}$. The $4$-cycle graph is of
higher degree in our construction if the group or the generators are
badly chosen, for instance if $\gG$ is Abelian, or if there are
commuting generators in $G_\omega$ or in $G_{\bar\omega}$ that are not
inverse of each other.
\end{remark}

\begin{remark} \rm
A closer look at the proof shows that $4$-cycles come from qubits
which are involved in two different ways in two generators.  It could
be thought that in order to avoid these $4$-cycles we just have to
ensure that each qubit is involved in the same way in all the
generators it belongs to. This is a bad solution. Assume for instance
that a qubit is involved in all its generators with label
$\omega$. Then the single qubit error with a $\omega$ at this position
and $0$ elsewhere generally belongs to the set of undetectable
errors. $4$-cycles are therefore unavoidable if we want stabilizer
codes with good error-correcting capabilities.
\end{remark}

\begin{example}\rm
We obtain a (6,12)-regular quantum LDPC code with the following
choices:
\begin{itemize}
\item $\gG=PSL_{2}(\mathbb{F}_{5})\times{PSL_{2}(\mathbb{F}_{5})}$
($PSL_2(\mathbb{F})$ denotes the quotient of group of $2 \times 2$
matrices over the field $\mathbb{F}$ of determinant $1$ by its center,
that is $\{\mathbb{I},-\mathbb{I}\}$;
\item $H = \{\I\}$ and $K=\{\I,u\}$ where $u^2 = \I$, with $u$ given
by $$u=\left(\begin{pmatrix} 1 & 2\\-1 & -1 \\\end{pmatrix}, \
  \begin{pmatrix} 1 & 2\\ -1 & -1 \\
\end{pmatrix}
\right);$$

\item $G_\omega\!=\left \{
\left(
\!\begin{pmatrix}
1 & 1 \\
0 & 1
\end{pmatrix}
,\mathbb{I}
\!\right)
,\!\ 
\left(
\!\begin{pmatrix}
0& -1 \\
1 & 0
\end{pmatrix}
,\mathbb{I}
\!\right)
,\!\ 
\left(
\!\begin{pmatrix}
1 & -1 \\
0 & 1
\end{pmatrix}
,\mathbb{I}
\!\right)
\right \}
$;
\item $
G_{\bar\omega}\!=\left \{
\left(
\mathbb{I}
,\!\ 
\begin{pmatrix}
1 & 1 \\
0 & 1
\end{pmatrix}
\!\right)
,\!\ 
\left(
\mathbb{I}
,\!\ 
\begin{pmatrix}
0& -1 \\
1 & 0
\end{pmatrix}
\!\right)
,\!\ 
\left(
\mathbb{I}
,\!\ 
\begin{pmatrix}
1 & -1 \\
0 & 1
\end{pmatrix}
\!\right)
\right \}
.$
\end{itemize}

This code has block length $n=3600$ and $n-k=1800$ generators. By
reducing the parity-check matrix in its standard form, see
\cite{Got97a}, we know that these generators are independent.  The
dimension of the code (that is the number of encoded qubits) is
therefore 1800. The code is of rate $1/2$.
\end{example}

\subsection{Construction of a (4,8)-regular quantum LDPC code}
\label{sec:4_8}

If we want to obtain a $(4,8)$-regular quantum LDPC code from the
construction given previously, it turns out that we cannot avoid
choosing an Abelian group for $\gG$. This would not be a wise choice as
it would induce many unwanted $4$-cycles. Here, we give an alternative
construction for $(4,8)$ codes such that the associated $4$-cycle
graph is $4$-regular. In addition, we impose that each column of $M$
has two entries with a $\omega$ and two with a $\bar\omega$. This defines a
quantum LDPC code of type $([2_\omega,2_{\bar\omega}],8)$.

Since the $4$-cycle graph will be constructed as a Cayley graph, we
remind that:
\begin{definition}
For $\gG$ a finite group and $S\subset \gG$ satisfying $S=S^{-1}$, the
Cayley graph $X(\gG,S)$ is the graph having $\gG$ as vertex set and
edges formed by pairs $\{x, y\}$ such that $x=gy$. $X(\gG,S)$ is
$k$-regular with $k=|S|$.
\end{definition}

Our alternative construction is obtained by following a $3$-step
procedure.

\paragraph{Step 1: construction of the $4$-cycle graph.}
The intuition behind our construction is the following remark which
explains how the $4$-cycle graph can be used in order to build the
Tanner graph. We notice that the subgraph of the $4$-cycle graph of a
$([2_\omega,2_{\bar\omega}],8)$ code induced by the qubits involved in
a generator is a $2$-regular graph with $8$ vertices. It is therefore
either a cycle of length $8$ or a union of cycles. We choose to
construct $([2_\omega,2_{\bar\omega}],8)$ codes for which all these
subgraphs are single cycles. Therefore, following certain cycles in
the $4$-cycle graph will reveal which qubits are involved in the
associated generator.

We obtain the $4$-cycle graph in the following way.  For $p$ prime
such that $p \equiv{1} \  \mod 4$, let $\mathbb{F}_p$ be the finite field
of size $p$ and consider
$$\gG = \{\ M\in{GL_{2}(\mathbb{F}_{p})}\ |\ (\det\ M)^{2}=\pm{1}\ \}.$$

Choose $g_+,\ g_- \in{GL_{2}(\mathbb{F}_{p})}$, such that:
\begin{itemize}
\item $\det g_+ = \pm 1$ and $\det g_- = \pm i$
\item $\gG$ is generated by the set
\begin{equation*}
S = \{g_+,\ g_+^{-1},\ g_-,\ g_-^{-1}\},
\end{equation*}
satisfying $(g_+g_-^{-1})^4=(g_-g_+)^4=\mathbb{I}$.
\end{itemize}
The graph $X(\gG, S)$ is {$4$-regular} and has $4p(p^2-1)$
vertices. This defines the $4$-cycle graph.

\paragraph{Step 2: Construction of a $(4,8)$-regular graph.}
It is easy to check that any qubit is involved in four cycles of
length $8$ corresponding respectively to the relations
$(g_+g_-)^4=\I,(g_-g_+^{-1})^4=\I,(g_-g_+)^4=\I,(g_-^{-1}g_+)^4=\I$. This
gives the $4$ generators in which this qubit is involved. The
$(4,8)$-regular graph is then obtained by putting an edge between each
qubit and the generators it belongs to.

\paragraph{Step 3: Edge labeling.} 
For the $(4,8)$-regular graph constructed in the previous step to be
the Tanner graph associated to a quantum LDPC code, we must now label
the edges in a way that commutation relations are satisfied. This can
be done in many different ways.  We give one way of performing this
task which is the one used for the example considered in the next
section.

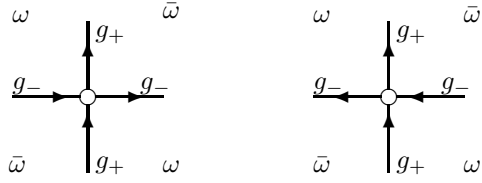
\begin{figure}
\begin{center}
\setlength{\unitlength}{2072sp}%
\begingroup\makeatletter\ifx\SetFigFont\undefined%
\gdef\SetFigFont#1#2#3#4#5{%
  \reset@font\fontsize{#1}{#2pt}%
  \fontfamily{#3}\fontseries{#4}\fontshape{#5}%
  \selectfont}%
\fi\endgroup%
\begin{picture}(5675,2010)(1306,-2356)
\put(1351,-1321){\makebox(0,0)[lb]{\smash{{\SetFigFont{10}{12.0}{\familydefault}{\mddefault}{\updefault}{\color[rgb]{0,0,0}$g_-$}%
}}}}
{\color[rgb]{0,0,0}\thinlines
\put(2251,-1411){\circle{180}}
}%
\thicklines
{\color[rgb]{0,0,0}\put(5851,-1321){\vector( 0, 1){585}}
}%
{\color[rgb]{0,0,0}\put(5851,-2311){\vector( 0, 1){720}}
}%
{\color[rgb]{0,0,0}\put(5851,-2311){\line( 0, 1){810}}
}%
{\color[rgb]{0,0,0}\put(5851,-1321){\line( 0, 1){810}}
}%
{\color[rgb]{0,0,0}\put(5761,-1411){\vector(-1, 0){585}}
}%
{\color[rgb]{0,0,0}\put(6751,-1411){\vector(-1, 0){675}}
}%
{\color[rgb]{0,0,0}\put(6751,-1411){\line(-1, 0){810}}
}%
{\color[rgb]{0,0,0}\put(5761,-1411){\line(-1, 0){810}}
}%
{\color[rgb]{0,0,0}\put(2251,-1321){\vector( 0, 1){585}}
}%
{\color[rgb]{0,0,0}\put(2341,-1411){\vector( 1, 0){585}}
}%
{\color[rgb]{0,0,0}\put(2251,-2311){\vector( 0, 1){720}}
}%
{\color[rgb]{0,0,0}\put(1351,-1411){\vector( 1, 0){675}}
}%
{\color[rgb]{0,0,0}\put(2251,-2311){\line( 0, 1){810}}
}%
{\color[rgb]{0,0,0}\put(1351,-1411){\line( 1, 0){810}}
}%
{\color[rgb]{0,0,0}\put(2341,-1411){\line( 1, 0){810}}
}%
{\color[rgb]{0,0,0}\put(2251,-1321){\line( 0, 1){810}}
}%
\put(5941,-691){\makebox(0,0)[lb]{\smash{{\SetFigFont{10}{12.0}{\familydefault}{\mddefault}{\updefault}{\color[rgb]{0,0,0}$g_+$}%
}}}}
\put(6481,-1321){\makebox(0,0)[lb]{\smash{{\SetFigFont{10}{12.0}{\familydefault}{\mddefault}{\updefault}{\color[rgb]{0,0,0}$g_-$}%
}}}}
\put(5941,-2266){\makebox(0,0)[lb]{\smash{{\SetFigFont{10}{12.0}{\familydefault}{\mddefault}{\updefault}{\color[rgb]{0,0,0}$g_+$}%
}}}}
\put(4951,-1321){\makebox(0,0)[lb]{\smash{{\SetFigFont{10}{12.0}{\familydefault}{\mddefault}{\updefault}{\color[rgb]{0,0,0}$g_-$}%
}}}}
\put(1351,-511){\makebox(0,0)[lb]{\smash{{\SetFigFont{10}{12.0}{\familydefault}{\mddefault}{\updefault}{\color[rgb]{0,0,0}$\omega$}%
}}}}
\put(3151,-2311){\makebox(0,0)[lb]{\smash{{\SetFigFont{10}{12.0}{\familydefault}{\mddefault}{\updefault}{\color[rgb]{0,0,0}$\omega$}%
}}}}
\put(3151,-466){\makebox(0,0)[lb]{\smash{{\SetFigFont{10}{12.0}{\familydefault}{\mddefault}{\updefault}{\color[rgb]{0,0,0}$\bar\omega$}%
}}}}
\put(1306,-2311){\makebox(0,0)[lb]{\smash{{\SetFigFont{10}{12.0}{\familydefault}{\mddefault}{\updefault}{\color[rgb]{0,0,0}$\bar\omega$}%
}}}}
\put(6751,-511){\makebox(0,0)[lb]{\smash{{\SetFigFont{10}{12.0}{\familydefault}{\mddefault}{\updefault}{\color[rgb]{0,0,0}$\bar\omega$}%
}}}}
\put(4951,-2311){\makebox(0,0)[lb]{\smash{{\SetFigFont{10}{12.0}{\familydefault}{\mddefault}{\updefault}{\color[rgb]{0,0,0}$\bar\omega$}%
}}}}
\put(4951,-511){\makebox(0,0)[lb]{\smash{{\SetFigFont{10}{12.0}{\familydefault}{\mddefault}{\updefault}{\color[rgb]{0,0,0}$\omega$}%
}}}}
\put(6751,-2311){\makebox(0,0)[lb]{\smash{{\SetFigFont{10}{12.0}{\familydefault}{\mddefault}{\updefault}{\color[rgb]{0,0,0}$\omega$}%
}}}}
\put(2341,-691){\makebox(0,0)[lb]{\smash{{\SetFigFont{10}{12.0}{\familydefault}{\mddefault}{\updefault}{\color[rgb]{0,0,0}$g_+$}%
}}}}
\put(2881,-1321){\makebox(0,0)[lb]{\smash{{\SetFigFont{10}{12.0}{\familydefault}{\mddefault}{\updefault}{\color[rgb]{0,0,0}$g_-$}%
}}}}
\put(2341,-2266){\makebox(0,0)[lb]{\smash{{\SetFigFont{10}{12.0}{\familydefault}{\mddefault}{\updefault}{\color[rgb]{0,0,0}$g_+$}%
}}}}
{\color[rgb]{0,0,0}\thinlines
\put(5851,-1411){\circle{180}}
}%
\end{picture}%

\end{center}
\caption{Subgraphs of the $4$-cycle graph showing the two types of
qubits encountered. Each quadrant corresponds to an $8$-cycle
associated with a generator in which the qubit has a non-zero
entry. The $\omega$'s and $\bar\omega$'s of each quadrant represent
the value of the entry for the corresponding generator obtained in the
labeling step.}\label{fig:bipartite}
\end{figure}

Let us first notice there is a natural bipartition of $\gG$, and hence
of the qubits, in two classes $\{g \in \gG, \det g = \pm 1\}$ and $\{g
\in \gG, \det g = \pm i\}$. The edges leaving a qubit of the first
class are labeled as follows. The two edges linking the qubit to the
generators corresponding to the cycles $(g_+g_-)^4 = \I$ and
$(g_-g_+)^4=\I$ get the label $\omega$; the two other edges receive
the label $\bar\omega$. The edges leaving the qubits of the second
class are labeled in the opposite way: the two edges linking such
qubit to the generators of type $(g_+g_-)^4 = \I$ and $(g_-g_+)^4=\I$
get the label $\bar\omega$, whereas the two other edges receive the
label $\omega$ (see Fig.\ref{fig:bipartite}).

It is then straightforward to check that this labeling defines
commuting generators.

\begin{remark}\rm
It should be noted that the $(4,8)$-regular quantum LDPC code obtained
with this construction looks in many ways like a (classical) cycle
graph. In particular the minimum weight of an undetectable error (in
other words the minimum distance of the quantum code) is at most
logarithmic in the number of qubits. This can be seen by considering
the subgraph of the Tanner graph consisting in taking only edges with
label $\omega$. This graph is $(2,4)$-regular. It has therefore a
cycle of logarithmic size in the number of vertices. It is
straightforward to check that the error consisting in putting a
$\omega$ at qubits belonging to this cycle and $0$'s elsewhere is an
undetectable error.
\end{remark}

\begin{example}\rm
We obtain a $(4,8)$ quantum LDPC code with the following choices:
\begin{itemize}
\item $p=13$;
\item $g_+=\begin{pmatrix}9&9\\12&10\end{pmatrix}$, $g_-=\begin{pmatrix}11&7\\5&6\end{pmatrix}$.
\end{itemize}
It can be checked readily that $n=8736$ and $k=4370$.
\end{example}
 
\section{Results}
\label{sec:results}

\begin{figure}
\begin{center}
\includegraphics[width=8cm]{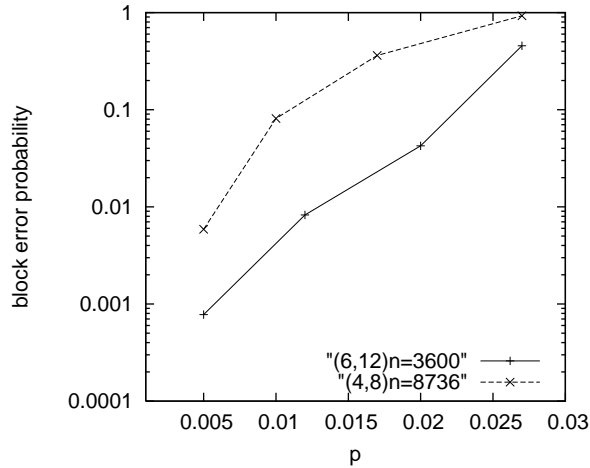}
\end{center}
\caption{Numerical simulations of the performance of a $(4,8)$ and a
$(6,12)$-regular quantum LDPC codes. The error estimation algorithm is
the syndrome version of MIN-SUM. The error model is the depolarizing
channel which acts on one-qubit density matrices as: $\rho \rightarrow
(1-p)\rho + \frac{p}{3}\left(X\rho X + Y\rho Y + Z\rho
Z\right)$.}\label{fig:simul}
\end{figure}

We present in Fig.~\ref{fig:simul} simulation results of two quantum
LDPC codes of rate $1/2$ on the depolarizing channel. We plot the
block error probability against the probability that there is an error
of type $X$,$Y$ or $Z$. The first code is a $(4,8)$-regular LDPC code
of length $8736$ and the second one is a $(6,12)$-regular LDPC code of
length $3600$. Despite the fact that the latter code is smaller, it
behaves much better with respect to the iterative error estimation
algorithm.

\section{Conclusion}
One of the drawbacks of the quantum LDPC codes constructed in this
article is that they have many cycles of size $4$. However, it should
be emphasized that this is not due to the particular construction
chosen here, but that it is a characteristic of all stabilizer codes.
This affects iterative decoding performances when the usual SUM-PRODUCT
or MIN-SUM algorithm is used. It would be interesting to study how
variants of this algorithm (see \cite{YFW04a} for instance) would
overcome this problem. In light of the fact that well chosen
(classical) irregular LDPC codes perform much better than their
regular counterpart, it would also be interesting to study whether our
construction could be generalized to yield irregular LDPC codes.

Finally, we would like to point out that quantum LDPC codes might be
good candidates for constructing fault-tolerant architectures. First,
we hope that the minimum distance of these codes increases linearly
with the block size, as it is the case for most classical LDPC
codes. This would warrant that any finite weight error can be
corrected for sufficiently large block sizes. Second, the rate of such
codes does not decrease to zero, thus possibly improving the overhead
requirements over schemes employing concatenation or toric
codes~\cite{DKLP02a}. Finally, and by definition, measurements
involved in the determination of the syndrome for quantum LDPC codes
is of fixed gate complexity, making it less prone to produce erroneous
syndromes than concatenated codes. If these are potential advantages
of using quantum LDPC codes for fault-tolerance, the issue of error
propagation in a circuit manipulating encoded information needs to be
addressed for them to become of practical interest. We chose to
postpone discussion of these issues to forthcoming publications.

This work was supported in part by ACI S\'ecurit\'e Informatique ---
R\'eseaux Quantiques. H.O. would like to thank D.\@~Poulin for
stimulating discussions and comments on an earlier version of the
manuscript.


\end{document}